%
\magnification=1200
\tolerance=1400
\overfullrule=0pt
\baselineskip=11.5pt

\font\rmc=cmr9
 1
\font\rma=cmbx9 scaled \magstep 2
\font\rmm=cmbx9 scaled \magstep 3
 2
 2

\font\tenib=cmmib10
\font\sevenib=cmmib10 at 7pt
\font\fiveib=cmmib10 at 5pt
\newfam\mitbfam
\textfont\mitbfam=\tenib
\scriptfont\mitbfam=\sevenib
\scriptscriptfont\mitbfam=\fiveib

\mathchardef\bfo="0\the\mitbfam21
\mathchardef\om"0\the\bffam0A
\font\bigastfont=cmr10 scaled \magstep 3
\def\bdot{\hbox{\bigastfont .}}
\def\ueber#1#2{{\setbox0=\hbox{$#1$}%
  \setbox1=\hbox to\wd0{\hss$\scriptscriptstyle #2$\hss}%
  \offinterlineskip
  \vbox{\box1\kern0.4mm\box0}}{}}

\def\R{\rm I\kern-.18em R}
\def\E{\rm I\kern-.18em E}
\def\ref{\par\noindent\hangindent\parindent\hangafter1}
\def\1{_{\vert}}

\topskip 1 true cm

\centerline{\rmm On Average Properties of}
\bigskip
\centerline{\rmm Inhomogeneous Fluids in General Relativity:}
\bigskip
\centerline{\rmm Perfect Fluid Cosmologies}

\bigskip
\bigskip
\centerline{Thomas Buchert}
\bigskip
\centerline{Universit\'e de Gen\`eve, D\'epartement de Physique Th\'eorique}
\smallskip
\centerline{24, quai E.~Ansermet, CH--1211 Gen\`eve 4, Switzerland}

\smallskip

\centerline{e--mail: buchert@theorie.physik.uni-muenchen.de}

\medskip\medskip

\centerline{\it G.R.G., accepted}

\vskip 1 true cm
\noindent
{\rmc
{\narrower
{\rma Summary:}
For general relativistic spacetimes filled with an irrotational perfect fluid 
a generalized form of Friedmann's equations governing the expansion factor of 
spatially averaged portions of inhomogeneous cosmologies is derived.
The averaging problem for scalar quantities is condensed into the problem of 
finding an `effective equation of state' including kinematical as well as 
dynamical `backreaction' terms that measure the departure from a standard 
FLRW cosmology. Applications of the averaged models are outlined including 
radiation--dominated and scalar field cosmologies (inflationary and 
dilaton/string cosmologies). In particular, the averaged equations show that 
the averaged scalar curvature must generically change in the course of 
structure formation, that an averaged inhomogeneous radiation cosmos does
not follow the evolution of the standard homogeneous--isotropic model, 
and that an averaged inhomogeneous perfect fluid features kinematical 
`backreaction' terms that, in some cases, act like a free scalar field source. 
The free scalar field (dilaton) itself, modelled by a `stiff' fluid, is 
singled out as a special inhomogeneous case where the averaged equations 
assume a simple form.

}}
 
\bigskip\bigskip\bigskip\bigskip
\noindent{\rmm 1. Introduction}
\bigskip\bigskip
\noindent
The present paper continues a line of research on average properties of
inhomogeneous fluids in general relativity that is based on a simple and 
intuitive averaging procedure. The simplification is guided by the 
restriction to scalar dynamical variables and to standard volume integration, 
in which case averaging is straightforward (for a discussion of 
alternative procedures see, e.g., Stoeger et al. 1999).   
Averaging is aimed at the construction of an {\it effective dynamics}
of spatial portions of the Universe from which, in principle, observable
average characteristics can be inferred like Hubble's constant, the 
effective 3--Ricci scalar curvature and the mean density of a given spatial domain,
which is bounded by the limits of observation. Naturally, this view entails a 
scale--dependent description of inhomogeneous cosmologies. 
In the case where the extension of the (simply--connected) spatial domain 
to the whole Universe is possible, such a description may allow 
to draw conclusions about global properties of the world models.

\smallskip

Paper I (Buchert 2000) 
was concerned with `dust cosmologies' restricting attention to the 
most popular inhomogeneous cosmologies. It is, however, desirable to extend
the regime of application of an `effective' (i.e. averaged) dynamics to a wider
range of spatial and temporal scales than that covered by the matter model 
`dust'. This is the motivation of the present work which presents the 
results for a large class of perfect fluid cosmologies. 

\smallskip

This class 
opens quite a piece of new terrain: it covers radiation--dominated cosmologies,
scalar field cosmologies including inhomogeneous dilaton/string cosmologies and 
inflationary cosmologies. It also extends the range of validity concerning 
averages of large--scale structure formation models for collisionless matter,
in which case the presence of a pressure--force that counteracts gravity is
implied by the development of multi--streaming within high--density regions;
here, it provides a phenomenological extension by including physics on
smaller spatial scales for the evolution of  
structure (see: Buchert \& Dom\'\i nguez 1998,
Buchert et al. 1999 in Newtonian cosmology; Maartens et al. 1999 in GR).

\bigskip
This paper is organized as follows. Sect.~2 presents Einstein's equations
for irrotational perfect fluids 
with the choice of foliation into flow--orthogonal hypersurfaces.
Averaging the scalar parts of Einstein's equations is investigated in Sect.~3.
The result is presented in a {\it Theorem} in Subsect.~3.2., which 
shows that the average expansion of inhomogeneous models is controlled by
`kinematical backreaction' due to shear and expansion fluctuations, and by
`dynamical backreaction' due to a non--vanishing pressure gradient in the 
hypersurfaces. It is manifest that the simple relation
between averaged 3--Ricci scalar curvature and `kinematical backreaction', as found for the 
matter model `dust' in Paper I (Buchert 2000), is supplemented by several
additional effects: besides dynamical contributions to `backreaction' 
the averaged energy-- and momentum conservation laws 
do not yield conservation laws for the averaged fields. This leads to 
more drastic changes in the average flow compared with the standard model.
{\it Corollary 1} and {\it Corollary 2} present compact formulations of the
averaged equations for {\it effective} sources. The relations between
additional sources in the generalized Friedmann equations are so formally
reduced to
the search for an {\it effective equation of state}.
While Sect.~3 appears rather formal, especially because the presence of 
pressure involves an inhomogeneous lapse function and so impairs the simplicity
of the equations, emphasis is focused on the application side in Sect.~4.
There we discuss some
relevant subcases that are members of the family of barotropic fluids: 
averaged `dust' models
are recovered from the more general framework, averaged 
radiation--dominated models display deviations from a standard radiation
cosmos, even if `kinematical backreaction' is absent, and the application to
inhomogeneous scalar field cosmologies is outlined.

\vfill\eject

\topskip 0 true cm
  
\noindent{\rmm 2. Einstein's Equations for Perfect Fluids}
\bigskip\bigskip
\noindent   
\noindent
{\rma 2.1. Choice of Foliation and Dynamical Variables}
\bigskip\medskip\noindent
We shall assume for the cosmic fluid 
that it is perfect and irrotational, so that we can introduce a foliation of spacetime 
into hypersurfaces orthogonal to the 4--velocity. It is not a problem to 
allow for a `tilted' slicing in order to include, e.g., vorticity (see, e.g.,
MacCallum \& Taub (1972), King \& Ellis (1973), 
Hwang \& Vishniac (1990), and Ellis et al. (1990)). 
For the applications we have in mind and also to keep the present
investigation transparent, we shall evaluate everything for this class of fluids.

For the purpose of averaging we shall consider a compact and simply--connected
domain contained within spatial hypersurfaces that are specified below.
This domain will be followed along the flow lines of the fluid elements;
thus we require that the total restmass of the fluid within the domain be
conserved. 

Let us first consider the (conserved) restmass flux 
vector$^1$\footnote{}{$^1$Greek indices run through
$0 ... 3$, while latin indices run through $1 ... 3$; 
summation over repeated indices is understood. A semicolon will denote 
covariant derivative with respect to the 4--metric with signature $(-,+,+,+)$; 
the units are such that $c=1$.}
$$
M^{\mu}: = \varrho u^{\mu}\;\;\;;\;\;\;M^{\mu}_{\;\,;\mu}=0\;\;\;
;\;\;\;\varrho > 0\;\;\;,\eqno(1a)
$$ 
where $\varrho$ is the restmass density and the flow lines are integral
curves of the 4--velocity $u^{\mu}$. 
Confining ourselves to irrotational fluids  
guarantees the existence of a scalar function $S$, such that
$$
u^{\mu} =: {-\partial^{\mu}S\over h}\;\;\;,
\eqno(1b)
$$
where the function $h$ will be identified below. It normalizes the 4--gradient 
$\partial^{\mu}S$ so that $u^{\mu}u_{\mu} = -1$,
$$
h = \sqrt{-\partial^{\alpha}S \partial_{\alpha}S} = 
u^{\mu} \partial_{\mu}S = 
:{\dot S}\;>\;0\;\;\;.\eqno(1c)
$$
The overdot stands for the material derivative operator along the flow lines
of any tensor field $\cal F$ as defined covariantly by
$$
\dot{\cal F}: = u^{\mu}{\cal F}_{\,\;;\mu} \;\;\;.  
\eqno(1d)
$$
\bigskip
We shall aim at a covariant description of the fluid flow with respect to
the natural foliation of spacetime into hypersurfaces {$S=const.$} 
representing the 3--dimensional `wave fronts' (for the covariant fluid 
approach compare Ellis \& Bruni 1989, Bruni et al. 1990a,b, Dunsby et al. 1992). 
With our choice of the fluid's 4--velocity (1b) we have to assure that
it remains time--like and, hence, the hypersurfaces 
$S = const.$ space--like. For this to be true 
the 4--gradient of the scalar field has to be time--like, 
$$
 \partial_{\alpha}S\partial^{\alpha}S = - h^2 < 0 \;\;\;.\eqno(1g)
$$
(For $h \in \R$ this is always true.)
As already noted, the definition (1b) implies that $u^{\mu}$ is irrotational 
(2a); it also implies 
that the covariant spatial gradient of $S$ in the hypersurfaces of constant $S$, 
denoted by $S_{||\mu}$, vanishes (2b), 
$$
\omega_{\mu\nu} = h_{\mu}^{\,\;\alpha}h_{\nu}^{\,\;\beta}u_{[\alpha;\beta]} 
= -h_{\mu}^{\,\;\alpha}h_{\nu}^{\,\;\beta}
({1\over h}\partial_{[\alpha}S)_{;\beta]} = 0
\;\;;\eqno(2a)
$$
$$
S_{||\mu} = h_{\mu}^{\;\,\nu}\partial_{\nu}S
= \partial_{\mu}S + u_{\mu}{\dot S}\; =\; 0 \;\;,\eqno(2b)
$$
where $h_{\mu\nu}: = g_{\mu\nu} + u_{\mu}u_{\nu}$ is the projection tensor into the hypersurfaces
{$S = const.$} orthogonal to the integral curves of the $4-$velocity $u^{\mu}$, 
$h_{\mu\nu}u^{\nu} = 0$.

Hence, $S(t)$ and $h (X^i ,t)$ play the role of `phase' and 
`amplitude' of the fluid's `wave fronts', 
respectively. 

\medskip\noindent
On these hypersurfaces we introduce the 3--metric $g_{ij}$
(the first fundamental form) that is induced by the 
projection, as well as
the extrinsic curvature tensor (the second fundamental form):
$$ 
h_{ij}:=g_{\mu\nu}h^{\mu}_{\;\,i}h^{\nu}_{\;\,j}=g_{ij}\;\;\;;\;\;\;
K_{ij}: = -u_{\mu;\nu}h^{\mu}_{\;\,i}h^{\nu}_{\;\,j} = -u_{i;j}\;\;.\eqno(3a,b)
$$
The final result will be covariant with respect to the given foliation,
but we shall label the flow lines by introducing (intrinsic) Gaussian
coordinates $X^i$ that appear in the line--element 
$$
ds^2 = -N^2 dt^2 + g_{ij}dX^i dX^j \;\;\;.\eqno(4a)
$$ 
Since by this choice of coordinates 
the velocities in 3--space vanish we are entitled to call $X^i$ Lagrangian coordinates.
In the language of the ADM formalism, which is put into perspective in the Appendix, we
have a vanishing shift vector in the hypersurfaces and the lapse function (together
with the 3--metric) encodes the inhomogeneities.
For scalar functions ${\cal F} = \psi$ the covariant derivative (1d) 
reduces to the total (or Lagrangian) derivative along the flow lines,
$$
{d\over d\tau} \psi := {d x^{\mu}\over dt}\partial_{\mu}\psi = 
u^{\mu}\partial_{\mu} \psi = {1\over N}\partial_t \psi\;\;\;,
\eqno(4b)
$$
where $N$ is the (inhomogeneous) lapse function. 
It is crucial to note that the latter operator corresponds to a total time derivative 
with respect to {\it proper time} 
$\tau$, which can be defined by $\tau : = \int{N dt}$.

\medskip\noindent
For later discussions we may express the symmetric tensor $K_{ij}$, or the expansion tensor $\Theta_{ij}:=-K_{ij}$,
respectively, in terms of kinematical quantities  
and their scalar invariants (Ehlers 1961). We decompose $\Theta_{ij}$
into its trace--free symmetric `shear tensor' $\sigma_{ij}:= 
\sigma_{\mu\nu}h^{\mu}_{\;\,i}h^{\nu}_{\;\,j}\;$,
$\sigma_{\mu\nu}u^{\nu} = 0$, 
and its trace, the `rate of expansion' $\theta: = u^{\alpha}_{\;\,;\alpha}$. 
From the decomposition $u_{\mu;\nu} = \sigma_{\mu\nu} + {1\over 3}\theta h_{\mu\nu} - {\dot u}_{\mu} u_{\nu}$
we have:  
$$
-K_{ij} = \Theta_{ij} = \sigma_{ij} + {1\over 3}\theta g_{ij}\;\;\;;\;\;\;-K = \theta 
\;\;\;.\eqno(5a)
$$
The tensor has three principal scalar invariants; in what follows we shall use
two of them:
$$
{\bf I}:= -K = \theta \;\;\;\;\;\;;\;\;\;\;\;2{\bf II}:= K^2 - K^i_{\,\;j} K^j_{\,\;i} 
= {2\over 3}\theta^2 - 2\sigma^2 \;\;\;,\eqno(5b,c)
$$
where we have introduced the `rate of shear' $\sigma$ by $\sigma^2 : = 
{1\over 2} \sigma^i_{\,\;j} \sigma^j_{\,\;i}$.

\bigskip\bigskip\bigskip

\noindent
{\rma 2.2. Basic Equations in 3 + 1 Form}
\bigskip\medskip\noindent
Einstein's equations for an irrotational perfect fluid with the energy--momentum
tensor
$$
T_{\mu\nu}: = \varepsilon u_{\mu}u_{\nu} + p h_{\mu\nu}  \;\;\;,
\eqno(6)
$$
with energy density $\varepsilon$ and pressure $p$,
may be cast into a set of
`constraint equations', the Hamiltonian and momentum 
constraints$^2$\footnote{}{$^2$
As before, a double vertical slash abbreviates the covariant derivative with respect to 
the 3--metric $g_{ij}$; for scalars it reduces to the partial derivative
with respect to Lagrangian coordinates denoted by a single vertical slash.}, 
$$\eqalignno{
{\cal R} + K^2 - K^i_{\,\;j} K^j_{\,\;i} &= 16\pi G \varepsilon \;\;\;,&(7a)\cr
K^i_{\,\;j || i} - K_{| j} &= 0\;\;\;,&(7b)\cr}
$$
and `evolution equations' for the the two fundamental
forms:
$$\eqalignno{
{d\over d\tau}{g}_{ij} &= -2 \;\,{g}_{ik}K^k_{\,\;j} \;\;,&(7c)\cr
{d\over d\tau}{K}^i_{\,\;j} &=  K K^i_{\,\;j} + {\cal R}^i_{\,\;j} - 
4\pi G \delta^i_{\,\;j}\left(\epsilon - p\right) 
- ( a^i_{\,\; || j} + a^i a_j )\;\;\;,&(7d)\cr}
$$ 
where the acceleration is completely contained in the hypersurfaces of 
constant $S$ and is defined as
$$
a_i = h^{\mu}_{\;\,i}a_{\mu}\;\;,\;\;a^{\mu}: = u^{\nu}u^{\mu}_{\,\;;\nu} = {\dot u}^{\mu} 
\;\;,\;\;a^{\mu}u_{\mu} = 0 \;\;,\eqno(8a)
$$
and ${\cal R}:= {\cal R}^i_{\,\;i}$, $K:=K^i_{\,\;i}$ 
denote the traces of the spatial Ricci tensor ${\cal R}^i_{\,\;j}$
and the extrinsic curvature tensor, respectively. 
Below, we shall only average the  
$4-$divergence $\cal A$ of the acceleration field:
$$
{\cal A}: = a^{\mu}_{\,\;;\mu} = a^i_{\,\; || i} + a^i a_i \;\;.\eqno(8b)
$$
A nonvanishing acceleration is a consequence of the fact that the pressure term
forces deviations from a geodesic flow.
From $T^{\mu\nu}_{\;\,\;\,;\nu} =0$ we 
derive the energy and momentum conservation laws:
$$
u_{\mu} T^{\mu\nu}_{\,\;\,\;;\nu} =0 \;\;\Leftrightarrow\;\; 
\dot\varepsilon = - \theta (\varepsilon + p) \;\;,\eqno(9a)
$$
$$
h_{\mu\alpha}T^{\mu\nu}_{\;\,\;\,;\nu}=0 \;\;\Leftrightarrow\;\; 
{\dot u}_{\alpha} = a_{\alpha} = -{1\over \varepsilon + p} \partial_{\mu} p 
h^{\mu}_{\;\,\alpha} = 
- {1\over \varepsilon + p} p_{| \alpha}\;\;\;.\eqno(9b) 
$$
Hence, 
$$
a_i = - {1\over \varepsilon + p} p_{|i} \;\;\;,\eqno(9c)
$$
and 
$$
{\cal A} =  - {1\over \varepsilon + p} p^{| i}_{\;\;||i} + 
{2\over (\varepsilon + p)^2}p^{| i}p_{| i} +
{1\over (\varepsilon + p)^2}p^{| i}\varepsilon_{| i}\;\;\;.\eqno(9d)
$$
From Eq. (1a) we also have the continuity equation for the restmass density
$$
\dot\varrho + \theta \varrho = 0\;\;\;.\eqno(9e)
$$
\medskip\noindent
According to (1c), $\dot S = {1\over N}\partial_t S(t) = h$, we 
can write Eq. (9d) completely in terms of the magnitude $h$ and its
spatial derivatives:
$$
{\cal A} = \left({N^{|i}\over N}\right)_{||i} = -{1\over h}h^{|i}_{\;\,||i}
+ {2\over h^2}h^{|i}h_{|i} \;=\; h\left({1\over h}\right)^{|i}_{\;\,||i}  
\;\;\;.\eqno(10a)
$$
Two other derived formulas will be used in what follows.
First, Raychaudhuri's equation, which follows by taking the trace of (7d) and 
inserting (7a):
$$
{\dot\theta} = -{1\over 3}\theta^2 - 2\sigma^2 - 4\pi G \left( \varepsilon + 3 p \right) + {\cal A}
\;\;\;,\eqno(10b) 
$$
and, second, an expression for the spatial Ricci curvature scalar
in terms of the energy source terms, the restmass density, 
the magnitude $h$ and its spatial derivatives: eliminating 
$2{\bf II} =  {2\over 3}\theta^2 - 2\sigma^2$ from Eq. (10b) and, using Eq. (5c), from
the Hamiltonian constraint Eq. (7a), we obtain with Eq. (9e): 
$$
{\cal R} = 12\pi G (\varepsilon - p) -\varrho {d^2 \over d\tau^2}
\left({1\over \varrho}\right)
+ h\left({1\over h}\right)^{|i}_{\;\,||i}  
\;\;\;.\eqno(10c) 
$$  
\bigskip\bigskip\bigskip

\noindent
{\rma 2.3. Thermodynamics of the Fluid}
\bigskip\medskip\noindent
First we note that the energy and restmass conservation laws Eqs. (9a,e)
are equivalent according to the first law of thermodynamics,
$$
{d\varrho \over \varrho} = {d\varepsilon\over \varepsilon + p}=:{ds\over s}
\;\;\;,\eqno(11a)
$$
upon dividing by $d\tau$. The latter equality defines the entropy density
$s$ that obeys the conservation law
$$
(s u^{\mu})_{;\mu} = 0 \;\;;\;\;\dot s + \theta s = 0\;\;\;.\eqno(11b)
$$
\smallskip\noindent
For closing the system of Einstein equations we need to identify a 
concrete matter model. 
Specific models are obtained by invoking a local `equation of state'  
relating the pressure with the other dynamical variables. 
We shall discuss examples that are all members of the class of 
`barotropic fluids', i.e., $p=\alpha(\varepsilon)$ is assumed to be locally
given and, in particular, the function $\alpha$ is the same for each 
fluid element (at each trajectory). 
The special inhomogeneous fluid cases discussed will all be contained in the simpler
class $\alpha (\varepsilon) = \gamma \varepsilon$ with $\gamma = const.$,
a `dust' matter model ($\gamma = 0$), a `radiation fluid' 
($\gamma = {1\over 3}$), and a `stiff fluid' corresponding to a free 
minimally coupled scalar field ($\gamma = 1$). 
\medskip
We shall now identify the normalization
amplitude $h$. Let us first derive $h$ for a `barotropic fluid'.
The momentum conservation law implies
$$
{N_{|i} \over N} = -{h_{|i} \over h} = -{p_{|i}\over \varepsilon + p} \;\;\;.\eqno(11b)
$$
Defining $\Pi : = \int {dp \over \varepsilon + p}$, with $\varepsilon =
\alpha^{-1}(p)$, we may write Eq. (11b) as
$(h_0\ln({h\over h_0})+ \Pi)_{|i}=0$, which may be integrated to give
$$
h \propto \exp{\Pi}\;\;\;,\eqno(11c)
$$ 
up to a time--dependent integration function.
Hence, we may write
$$
{dh \over h} = {dp \over \varepsilon + p}\;\;\;.\eqno(11d)
$$ 
In general, we identify the magnitude $h$ with the 
 `injection energy per fluid element and unit
restmass' (Israel 1976), 
$$
h := {\varepsilon + p \over \varrho}\;\;\;,\eqno(11e)
$$
which is related to the relativistic enthalpy 
$\eta: = {\varepsilon + p \over n}$ 
by $h = \eta/m$ with $m$ the unit restmass of a fluid element,
and $n$ the baryon density.
Note that Eq. (11d) holds by defining $h$ as in Eq. (11e) as a result of
the conservation laws Eqs. (9a,e), since from Eq. (11a),
$$
d\varepsilon = h d\varrho\;\;\;.\eqno(11f)
$$ 
For a barotropic fluid we can easily see that $\varepsilon$ is a function of the
restmass density only and, hence, $h$ is a function of $\varrho$.
The evolution equation for $h$ in this case (and in the simpler case
$p = \gamma\varepsilon$) reads:
$$
\dot h + \theta \alpha' (\varepsilon) h = \dot h  + \gamma\theta h = 0\;\;\;;\eqno(11g)
$$
$h$ obeys a simple continuity equation in the case of a `stiff' fluid 
with $\gamma = 1$.
\bigskip
The discussion of special cases will be resumed in Sect. 4.

\vfill\eject

\noindent
{\rmm  3. The Averaged System}      
\smallskip
\bigskip\bigskip\noindent
{\rma 3.1. The Averaging Procedure}
\bigskip\medskip\noindent
Spatially averaging equations for scalar fields 
is a covariant operation given a foliation of spacetime.
Therefore, we shall in what follows only consider scalar functions 
$\Upsilon (X^i ,t)$. We shall define
the averaging operation by the usual spatial volume average performed on 
an arbitrary compact support of the fluid ${\cal D}$ 
contained within the hypersurfaces
$S(t) =  const.$:
$$
\langle \Upsilon \rangle_{\cal D}: = 
{1\over V_{\cal D}}\int_{{\cal D}} \Upsilon \; J d^3 X 
 \;\;\;,\;\;\;J:=\sqrt{\det({g}_{ij})}\;\;\;.\eqno(12a)
$$
The volume of the region itself (set $\Upsilon = 1$) is given by
$V_{\cal D}(t) : = \int_{\cal D} J d^3 X$.

\noindent
We also introduce a dimensionless scale factor via the volume (normalized by the volume of the initial domain
${V_{\cal D}}_o$):
$$
a_{\cal D} (t) : = \left({V_{\cal D}\over {V_{\cal D}}_o}\right)^{1/3} \;\;\;.
\eqno(12b)
$$
This means that we are only interested in the effective dynamics of the domain; $a_{\cal D}$ will be a 
functional of the domain's shape (dictated by the metric) and position. Since the domains follow the 
flow lines, the total restmass 
$M_{\cal D}: = \int_{\cal D} \varrho J d^3 X $
contained in a given domain is conserved. 

\noindent
The following formulas are crucial for evaluating averages.
Taking the trace of Eq. (7c), written in the form 
$$
K^i_{\,\;j} = -{1\over 2}{g}^{ik}{d\over d\tau}{g}_{kj}\;\;\;,
$$
we obtain with ${1\over 2}{g}^{ik}{d\over d\tau}{g}_{ki}
= (\ln J)^{\bdot}$ the identity
$$
\dot J = \theta J \;\;\;.\eqno(12c)
$$
The rate of change of the volume $V(t)$ in the hypersurfaces $S(t)=const.$
is evaluated by taking the partial time derivative of the volume and dividing by
the volume.
Since $\partial_t$ and $d^3 X$ commute (but not ${d\over d\tau}$ 
and $d^3 X$ !) we obtain:
$$
{\partial_t V_{\cal D} (t)\over V_{\cal D} (t)} = {1\over V_{\cal D} (t)}
\int_{\cal D} \partial_t J d^3 X = {1\over V_{\cal D} (t)}
\int_{\cal D} N {\dot J} d^3 X = {1\over V_{\cal D} (t)}
\int_{\cal D} N\theta J d^3 X = \langle N\theta\rangle_{\cal D}\;\;.\eqno(12d) 
$$
Introducing the scaled (t--)expansion ${\tilde\theta}:=N\theta$ we define
an effective (t--)Hubble function in the hypersurfaces by
$$
\langle \tilde\theta \rangle_{\cal D} = {\partial_t V_{\cal D} (t) \over 
V_{\cal D} (t)} = 3 
{\partial_t a_{\cal D} \over a_{\cal D}} =: 3 {\tilde H}_{\cal D}\;\;\;.\eqno(12e)
$$
(Notice that we reserve the overdot for the covariant derivative.)
\bigskip\noindent 
It is now straightforward to prove the following {\it Lemma}  
for an arbitrary scalar field $\Upsilon (X^i ,t)$:

\bigskip\noindent
\underbar{\bf Lemma$\;\;$({\it Commutation rule})} 
\bigskip
$$
\partial_t \langle \Upsilon\rangle_{\cal D} - \langle{\partial_t \Upsilon}
\rangle_{\cal D} = \langle \Upsilon\tilde\theta\rangle_{\cal D} - 
\langle \Upsilon\rangle_{\cal D}\langle\tilde\theta\rangle_{\cal D}
\;\;\;,\eqno(12f)
$$
or, alternatively,
$$
\partial_t \langle\Upsilon\rangle_{\cal D} 
+ 3{\tilde H}_{\cal D}\langle \Upsilon\rangle_{\cal D}
= \langle\partial_t \Upsilon + \Upsilon{\tilde{\theta}}\rangle_{\cal D} \;\;\;.\eqno(12g)
$$
\bigskip
A simple application of this {\it Lemma} is the proof that the total restmass
in a domain is conserved: let $\Upsilon = \varrho$, then 
$\partial_t \langle \varrho\rangle_{\cal D}+3 {\tilde H}_{\cal  D}
\langle \varrho\rangle_{\cal D} = \langle {\partial_t \varrho}+ \varrho\tilde\theta
\rangle_{\cal D}=0$ according to the local conservation law Eq. (9e). {\bf q.e.d.}
\bigskip\bigskip
\noindent
{\rma 3.2. Averaged Equations for Irrotational Perfect Fluids}
\bigskip\medskip\noindent
Averaging Raychaudhuri's equation (10b) and 
the Hamiltonian constraint (7a) 
with the help of the prescribed procedure, 
we end up with the following two equations for the scale factor of the
averaging domain which we may formulate in the form of a theorem:
\bigskip\smallskip
\noindent
\underbar{\bf Theorem -- Part I $\;\;$({\it Equations for the effective scale factor})}
\bigskip\noindent
The spatially averaged equations for the scale factor $a_{\cal D}$, 
respecting restmass conservation, read:
\smallskip\noindent
averaged Raychaudhuri equation for the scaled (t--)densities 
${\tilde\varepsilon}:=N^2 \varepsilon$ and ${\tilde p}:=N^2 p$: 
$$
3{\partial_t^2 a_{\cal D} \over a_{\cal D}} + 
4\pi G \langle {\tilde\varepsilon} + 3{\tilde p}\rangle_{\cal D}
\;=\; {\tilde{\cal Q}}_{\cal D} + {\tilde{\cal P}}_{\cal D}
\;\;\;;\eqno(13a)
$$
averaged Hamiltonian constraint:
$$
6 {\tilde H}_{\cal D}^2 - 16\pi G \langle {\tilde\varepsilon} \rangle_{\cal D} 
\;=\; -\left({\tilde{\cal Q}}_{\cal D}
+ \langle {\tilde{\cal R}} \rangle_{\cal D}\right) \;\;\;,\eqno(13b)
$$
where we have introduced the scaled spatial (t--)Ricci scalar 
${\tilde{\cal R}}:= N^2 {\cal R}$, and we have separated off the 
domain dependent `backreaction' terms: the \underbar{\it kinematical backreaction}, 
$$
{\tilde{\cal Q}}_{\cal D} : = \lbrack 2 \langle N^2 {\bf II}\rangle_{\cal D} 
- {2\over 3}\langle N{\bf I}\rangle_{\cal D}^2 \rbrack\;=\;
{2\over 3}\langle\left({\tilde\theta} 
- \langle{\tilde\theta}\rangle_{\cal D}\right)^2 
\rangle_{\cal D} - 2\langle{\tilde\sigma}^2\rangle_{\cal D}
\;\;\;,\eqno(13c)
$$
with the scaled (t--)shear scalar ${\tilde\sigma}: = N\sigma$,
and the \underbar{\it dynamical backreaction},
$$
{\tilde{\cal P}}_{\cal D}: = \langle {\tilde{\cal A}}\rangle_{\cal D}
+ \langle {\dot N}{\tilde\theta}\rangle_{\cal D} \;\;\;,\eqno(13d)
$$
with the scaled (t--)acceleration divergence ${\tilde{\cal A}}:=N^2 {\cal A}$.

\medskip\noindent\underbar{Note:}$\;$
Eq. (13a) can also be obtained by an argument given by 
Yodzis (1974), which is summarized in Appendix C of (Paper I).
\bigskip\noindent
The  source terms on the r.--h.--s. of Eq.~(13a) include the 
`kinematical backreaction' (13c) that describes the impact of 
inhomogeneities on the scale factor due to averaged shear and expansion fluctuations. 
It vanishes for the standard FLRW cosmologies.
Additionally, Eq.~(13a) features another `dynamical backreaction' term
$\langle {\tilde{\cal A}} \rangle_{\cal D}$ together with 
a technical term due to the change of the lapse function.
The former term also vanishes for standard FLRW cosmologies; both terms
vanish for zero pressure.
Note that the averaged Hamiltonian constraint does not 
involve pressure terms as expected. 

\smallskip\noindent
These equations show that the averaged shear fluctuations tend to increase the
expansion rate similar to the effect of the averaged energy source terms 
(provided the energy condition $\langle{\tilde\varepsilon} + 3{\tilde p}\rangle_{\cal D} > 0$ holds), 
while the averaged expansion fluctuations work in the direction 
of stabilizing structures. Pressure
forces can do both; the sign of the averaged divergence of the 4--acceleration 
can be positive or negative.
In the Newtonian framework one can show that, to a first--order approximation, 
the combined effect of gravity
and pressure leads to stabilization of structures (Buchert \& Dom\'\i nguez 1998, 
Buchert et al. 1999, Adler \& Buchert 1999).
Note, however, that since pressure is a source of the gravitational field energy too, 
it is harder to oppose the gravitational collapse
than in the corresponding Newtonian treatment 
(compare the terms which add positive contributions in Eq. (9d)
with their Newtonian analogues). 
\medskip\noindent
We proceed by calculating the integrability condition for the system of 
equations (13a,b), i.e., we shall answer the question which equation has to 
hold in order that (13b) be the
integral of (13a). For this end we take the partial 
time--derivative of (13b) and insert
into the result again our starting set of equations (13a) and (13b). We get:
\bigskip\bigskip
\noindent
\underbar{\bf Theorem -- Part II $\;\;$({\it Integrability and energy
balance conditions})}
\smallskip
Eq.~(13b) is an integral of Eq.~(13a),  iff
$$
\partial_t {\tilde{\cal Q}}_{\cal D} + 6 {\tilde H}_{\cal D} {\tilde{\cal Q}}_{\cal D} + 
\partial_t \langle {\tilde{\cal R}} \rangle_{\cal D}
+ 2 {\tilde H}_{\cal D} \langle {\tilde{\cal R}} \rangle_{\cal D} + 
4 {\tilde H}_{\cal D} {\tilde{\cal P}}_{\cal D}
$$
$$
- 16\pi G \lbrack \partial_t \langle{\tilde\varepsilon} \rangle_{\cal D} 
+ 3 {\tilde H}_{\cal D}\langle {\tilde\varepsilon} + {\tilde p} \rangle_{\cal D}\rbrack 
 \;=\;0\;\;\;.\eqno(14a)
$$   
The expression involving the energy density and the pressure 
does not vanish in general. To see this we average the 
local energy conservation law (9a). We obtain:
$$
\partial_t \langle \varepsilon \rangle_{\cal D} + 3 {\tilde H}_{\cal D}\langle 
\varepsilon + p \rangle_{\cal D} = 
\langle \partial_t p \rangle_{\cal D} - \partial_t \langle p \rangle_{\cal D}
\;\;\;.
\eqno(14b)
$$
For the scaled (t--)variables we accordingly have for the local law:
$$
\partial_t {\tilde\varepsilon} + {\tilde\theta}({\tilde\varepsilon} + 
{\tilde p}) = 2 {\dot N}{\tilde\varepsilon}\;\;\;,\eqno(14c)
$$
and for the average:
$$
\partial_t \langle {\tilde\varepsilon} \rangle_{\cal D} + 3 {\tilde H}_{\cal D}\langle 
{\tilde\varepsilon} + {\tilde p} \rangle_{\cal D} = 
\langle \partial_t {\tilde p} \rangle_{\cal D} - \partial_t \langle {\tilde p} 
\rangle_{\cal D} + \langle 2 {\dot N} {\tilde\varepsilon}\rangle_{\cal D}
\;\;\;.\eqno(14d)
$$
\bigskip\bigskip\noindent
This shows that the pressure term introduces a possibly interesting effect.
In the `dust' case 
(Paper I: Corollary 1)
the averaged fields obey the same equations as the local fields provided we use their representation in terms
of invariants of the second fundamental form. Here, this is no longer true. 
In particular, Eq. (14b) shows that the averaged energy conservation law invokes
non--commuting terms that are nonzero for inhomogeneous fluids.
Thus, even if both the `kinematical' and `dynamical backreaction' terms are
assumed to be negligible or cancel for ``some'' reason, the
averaged model is different from the standard homogeneous--isotropic models.
This fact will be manifest in the following more compact alternative representations 
of the averaged equations. 
\bigskip\bigskip
\noindent
\underbar{\bf Corollary 1 $\;\;$({\it Averaged equations: first effective form})}
\bigskip\noindent
Let us define effective densities as follows:
$$
\eqalignno{
\varepsilon^{(1)}_{\rm eff}:=& \langle{\tilde\varepsilon}\rangle_{\cal D} - 
{{\tilde{\cal Q}}_{\cal D}\over 16\pi G} \;\;\;,&(15a)\cr
p^{(1)}_{\rm eff}:=& \langle {\tilde p} \rangle_{\cal D} - {{\tilde{\cal Q}}_{\cal D}\over 16\pi G} 
- {{\tilde{\cal P}}_{\cal D}\over 12\pi G}\;\;\;.&(15b)\cr} 
$$
Then, the averaged equations can be cast into a form similar to the standard Friedmann equations:
$$
3{{\partial_t^2 a}_{\cal D} \over a_{\cal D}} + 4\pi G 
\left(\varepsilon^{(1)}_{\rm eff} 
+ 3p^{(1)}_{\rm eff}\right)\;=\; 0\;\;\;;\eqno(15c)
$$
$$
6 {\tilde H}_{\cal D}^2 + \langle {\tilde{\cal R}} \rangle_{\cal D} 
- 16\pi G \varepsilon^{(1)}_{\rm eff} 
\;=\; 0\;\;\;,\eqno(15d)
$$
and the integrability condition of (15c) to yield (15d) has the form of a 
balance equation between the effective sources and the averaged spatial (t--)Ricci scalar:
$$
\partial_t \varepsilon^{(1)}_{\rm eff} + 3 {\tilde H}_{\cal D}
\left(\varepsilon^{(1)}_{\rm eff} + p^{(1)}_{\rm eff}\right)
 = {1\over 16\pi G} \left( \partial_t\langle {\tilde{\cal R}}\rangle_{\cal D} +
 2 {\tilde H}_{\cal D} \langle {\tilde{\cal R}}\rangle_{\cal D} 
\right)\;\;\;.\eqno(15e)
$$
The effective densities obey a conservation law, if the domains' curvature
evolves like in a ``small'' FLRW cosmology, 
$\langle {\tilde{\cal R}}\rangle_{\cal D}=0$, or 
$\langle {\tilde{\cal R}}\rangle_{\cal D}\propto a_{\cal D}^{-2}$,
respectively. In particular, Eq. (15e) shows that in general the averaged 
densities are directly coupled to the evolution of the averaged spatial curvature.

\bigskip
Considering the averaged spatial t--Ricci scalar as an effective source as well, 
one may cast the equations into an even more 
elegant form. 
\vfill\eject
\noindent
\underbar{\bf Corollary 2 $\;\;$({\it Averaged equations: second effective form})}
\bigskip\noindent
Defining
$$
\eqalignno{
\varepsilon^{(2)}_{\rm eff}:=& \langle{\tilde\varepsilon}\rangle_{\cal D} 
- {{\tilde{\cal Q}}_{\cal D}\over 16\pi G} 
- {\langle {\tilde{\cal R}}\rangle_{\cal D}\over 16\pi G}\;\;\;,&(16a)\cr
p^{(2)}_{\rm eff}:=& \langle {\tilde p} \rangle_{\cal D} - 
{{\tilde{\cal Q}}_{\cal D}\over 16\pi G} 
+ {\langle {\tilde{\cal R}}\rangle_{\cal D}\over 48\pi G} - 
{{\tilde{\cal P}}_{\cal D}\over 12\pi G}\;\;\;,&(16b)\cr} 
$$
we obtain equations that assume the form of spatially 3--Ricci flat Friedmann cosmologies:
$$
3{{\partial_t^2 a}_{\cal D} \over a_{\cal D}} + 
4\pi G \left(\varepsilon^{(2)}_{\rm eff} + 
3p^{(2)}_{\rm eff}\right)\;=\; 0\;\;\;;\eqno(16c)
$$
$$
6 {\tilde H}_{\cal D}^2 - 16\pi G \varepsilon^{(2)}_{\rm eff} 
\;=\; 0\;\;\;,\eqno(16d)
$$
and the integrability condition of (16c) to yield (16d) has exactly the form of a conservation law:
$$
\partial_t \varepsilon^{(2)}_{\rm eff} + 3 {\tilde H}_{\cal D}
\left(\varepsilon^{(2)}_{\rm eff} + p^{(2)}_{\rm eff} \right)
\; = \;0\;\;\;.\eqno(16e)
$$

\noindent
\underbar{\bf Remarks:}
\medskip\noindent
These alternative representations reduce the solution of the averaging problem
for scalars, at least formally, to the problem of finding
an `effective equation of state' that relates the effective densities. 
Relativistic Lagrangian perturbation schemes will be useful to establish such
relations. Looking at Eqs.~(15) it is interesting to note that  
the `kinematical backreaction' term itself effectively performs 
like a free scalar field source, or like 
`stiff matter' in the case $\langle{\tilde{\cal R}}\rangle_{\cal D} 
\propto a_{\cal D}^{-2}$. However, care must be taken with such
statements, since ${\tilde{\cal Q}}_{\cal D}$ is related to 
$\langle{\tilde{\cal R}}\rangle_{\cal D}$, and there is no a priori 
reason why 
$\langle{\tilde{\cal R}}\rangle_{\cal D} \propto a_{\cal D}^{-2}$ 
if backreaction is present.
It nevertheless suggests to separate off the `stiff component' from the 
`effective equation of state': we already noted that
$-{1\over 16\pi G}{\tilde{\cal Q}}_{\cal D}$ in {\it Corollary 1}
forms a `stiff' part; deviations
from `stiffness' are, apart from those due to the matter sources, due to the `dynamical
backreaction' $-{1\over 12\pi G}{\tilde{\cal P}}_{\cal D}$. In the form of 
{\it Corollary 2} the `stiff' part is $-{1\over 16\pi G}
({\tilde{\cal Q}}_{\cal D} +\langle {\tilde{\cal R}}\rangle_{\cal D})$, and the deviations
from `stiffness' are due to the term ${1\over 12\pi G}(\langle {\tilde{\cal R}}
\rangle_{\cal D} - {\tilde{\cal P}}_{\cal D})$. If `dynamical backreaction'
compensates the averaged scalar curvature, then the whole `backreaction' 
forms a `stiff component'. The condition for this can be inferred from Eqs.~(13a,b)
yielding the general relation:
$$
\langle {\tilde{\cal R}}
\rangle_{\cal D} - {\tilde{\cal P}}_{\cal D}=-3({{\partial_t^2 a}_{\cal D}
\over a_{\cal D}} + 2 {\tilde H}_{\cal D}^2 ) + 12\pi G\langle {\tilde\varepsilon} - {\tilde p} 
\rangle_{\cal D}\;\;\;.
\eqno(17)
$$  
\medskip
In general, the system of equations (13,14) is {\it not} a closed system, which can
be most easily seen in the form of {\it Corollary 2}: we have three equations
(16c,d,e) for the three variables $a_{\cal D}$, $\varepsilon^{(2)}_{\rm eff}$ and
$p^{(2)}_{\rm eff}$, but only two of them are independent.  
We need an effective equation of state to close the system.  

\vfill\eject
\noindent
{\rmm 4. Discussion of Subcases}
\bigskip\bigskip\noindent
Since this paper is meant to provide the basic architecture for 
applications, let us note the following useful formulas.
\medskip
Firstly, the equations of Sect.~3 simplify by using the following
reparametrization of time:
the line element is invariant under the change of the time coordinate
$t \mapsto S(t)$, so there is still some gauge freedom.
Using the `phase fronts' as the new time coordinate we define a new lapse
function $\tilde N$ by
$$
N dt = :{\tilde N} dS\;\;\;,\;\;\;{\rm i.e.}\;\;\;,\;\;\;
N = {\tilde N} \partial_t S
\;\;\;.\eqno(18a)
$$
In particular, the total (Lagrangian) derivative becomes
$$
{d\over d\tau} = {1\over N}\partial_t = {1\over {\tilde N}}\partial_S 
=h \partial_S\;\;\;,
\eqno(18b)
$$
where the latter equality follows from Eqs. (1c) and (4b). 

Notice that with this new choice of time coordinate all equations of
Sect.~3 remain form invariant, if $N$ is replaced by $\tilde N$, and
partial time--derivatives are replaced by partial derivatives with respect to
$S$. The latter will be abbreviated by a prime in what follows. All fields are functions of 
the independent variables $(X^i, S)$ now. 
\medskip

Secondly, for the evaluation of the terms appearing in the averaged equations it is
helpful to note the following simplifications.
We shall give expressions for a barotropic fluid and, especially, for the simple class
of matter models $p=\gamma \varepsilon$, which is relevant for many applications.

For barotropic fluids,
$$
\dot\varepsilon + \theta (\varepsilon + \alpha(\varepsilon)) = 0 \;\;\;,\eqno(18c)
$$
we can integrate the energy conservation law along trajectories of fluid elements
using $\dot J = \theta J$ (Eq. 12c) and $\dot\varrho = -\theta \varrho$ 
(Eq. 9e) to find the entropy density,
$$
s(\varepsilon) \propto J^{-1} \;\;\;\;{\rm with}\;\;\;\;s(\varepsilon) \propto 
\exp{\int {d\varepsilon \over \varepsilon + \alpha (\varepsilon)}}\;\;,\eqno(18d) 
$$
and, upon performing the integral, the energy density. 
In particular, for $\alpha' = \gamma = const.$ we obtain: 
$$
s(\varepsilon) \propto \varepsilon^{{1\over 1+\gamma}}\;\;\;\;;\;\;\;\;
\varepsilon \propto {1\over J^{1+\gamma}}\;\;\;.\eqno(18e)
$$
Hence, with $\varrho \propto J^{-1}$, and using Eq. (11g),
$$
\varepsilon \propto \varrho^{1+\gamma}\;\;\;\;;\;\;\;\;h 
\propto \varrho^{\gamma}\;\;\;\;;\;\;\;\;s\propto \varrho\;\;\;.\eqno(18f) 
$$
\medskip
Furthermore,  
the scaled (t--)variables are now normalized with respect to the magnitude $h$
or its square, respectively. E.g., we have
$$
{\tilde N} = {1\over h}\;\;\;;\;\;\;{\tilde\theta} = {\tilde N}\theta
= {\theta\over h}\;\;\;;\;\;\;{\tilde\varepsilon} = 
{\varepsilon\over h^2}
\;\;\;;\;\;\;{\rm etc.}\;\;\;.\eqno(18g)
$$
For barotropic fluids with $\alpha' = \gamma =const.$ the normalization function
can be written in powers of the restmass density$^3$\footnote{}{$^3$The constants
$C_1$ and $C_2$ appearing in the following equations are in general 
$X^i$--dependent; also equations of state, if they arise as integrals
of dynamical equations, involve $X^i$--dependent functions of integration
such as $\gamma$. One may use the freedom to relabel the fluid elements such that
the constants or some product of them are $X^i$--independent. We here assume
that all the constants are equal for each fluid element for notational ease;
it is straightforward to write down the more general expressions, if needed.}. 
Also for this case
the expression involving the covariant derivative of the lapse function in
Eq.~(13d) is simply proportional to the $t-$expansion rate (upon using the
integral for $h$ Eq.~(18f)):
$$
{\dot {\tilde N}}  = {{\tilde N}'\over {\tilde N}} = - {h' \over h} = -\gamma 
{\varrho' \over \varrho} = \gamma {\tilde\theta}
\;\;\;.\eqno(18h) 
$$
The term involving the change of the lapse function in Eq.~(14d)
is simply time--dependent in two cases: it can be written as follows 
for the homogeneous case (Eq.~(18i); Subsect.~4.1), and for the
`stiff fluid' representing a free minimally coupled scalar field 
(Eq.~(18j); Subsect.~4.4):
$$\eqalignno{
\langle 2{\dot {\tilde N}}{\tilde\varepsilon}\rangle_{\cal D} &= 
6{\tilde H}\gamma{\varepsilon_H \over h_H^2}\;\;\;,&(18i)\cr
\langle 2{\tilde\theta}\tilde\varepsilon\rangle_{\cal D} &=
3{\tilde H}_{\cal D}\;\;\;.&(18j)\cr}
$$ 
It is interesting that the latter (fully inhomogeneous) case implies 
simplifications such that, e.g., Eqs.~(14c,d) reduce to identities (since
in this case $\tilde{\varepsilon} = \tilde{p} = {1\over 2}$, compare 
Subsect.~4.3).
\medskip
For the `dynamical backreaction' we have for $p=\gamma \varepsilon$: 
$$
{\tilde{\cal P}}_{\cal D} = \langle {\tilde{\cal A}}\rangle_{\cal D} + 
\gamma \langle {\tilde\theta}^2
\rangle_{\cal D}\;\;\;,\eqno(18k)
$$
and, noting that ${\tilde\theta} = - {\varrho' \over \varrho}$,
and using Eqs.~(10a) and (18f) with $h=C_1 \varrho^{\gamma}$, 
the `dynamical backreaction' term can be entirely written in terms of the
restmass density:
$$
{\tilde{\cal P}}_{\cal D} = \langle {\tilde{\cal A}}\rangle_{\cal D} + \gamma 
\langle{\tilde\theta}^2 \rangle_{\cal D} = {\gamma\over C_1^2}
\langle -{1\over \varrho^{2\gamma+1}}\varrho^{|i}_{\;\,||i}
+ {(1+\gamma)\over \varrho^{2\gamma+2}}\varrho^{|i}\varrho_{|i} + 
C_1^2 {\varrho'^2 \over \varrho^2}\rangle_{\cal D}\;\;\;.\eqno(18l) 
$$
The same is true for the averaged spatial $t-$Ricci curvature using Eqs. (10c)
and (18f) with $\varepsilon = C_2 \varrho^{1+\gamma}$:
$$
\langle{\tilde{\cal R}}\rangle_{\cal D} = 12\pi G {C_2 \over C_1^2}
(1-\gamma)\langle \varrho^{1-\gamma}
\rangle_{\cal D} - \varrho \left({1\over \varrho}\right)^{''} + 
{\tilde{\cal P}}_{\cal D}\;\;\;,\eqno(18m)
$$
where from Eq. (11e) $C_1 = (1+\gamma)C_2$, and $C_2$ is determined by initial 
conditions.
This makes the problem accessible to relativistic Lagrangian perturbation
models, since the restmass density can be integrated exactly along the flow 
lines, $\varrho \propto J^{-1}$, and $J$ can be computed from the basic
dynamical variable in Lagrangian perturbation theory (see, e.g.,
Kasai 1995, Takada \& Futamase 1999 for relativistic `dust' models, 
Adler \& Buchert 1999 for pressure--supported fluids in Newtonian theory).

\vfill\eject

\noindent
{\rma 4.1. Homogeneous--isotropic Cosmologies}
\bigskip\medskip\noindent
The requirement of homogeneity and isotropy
reduces Eqs. (13a,b) to the familiar Friedmann 
equations. We may simply put the lapse function (without loss of generality)
equal to $1$ in Eqs. (13a,b) 
and find ${\tilde{\cal Q}}_{\cal D}={\tilde{\cal P}}_{\cal D}=0$, and:
$$
3{{\ddot a} \over a} + 4\pi G (\varepsilon_H + 3p_H )\;=\; 0
\;\;\;;\eqno(19a)
$$
$$
6 H^2 + {\cal R}_H  - 16\pi G \varepsilon_H  
\;=\; 0\;\;\;;\;\;\;H:={{\dot a} \over a}\;\;\;.\eqno(19b)
$$ 
Already before averaging $\varepsilon_H$ and $p_H$ are functions of time only,  
and the scale factor assumes its global 
standard value $a_{\cal D} \equiv a$. The domain dependence has disappeared.
Note also that the integrability condition (14a) reduces to the equation for the 
3--Ricci curvature scalar of the spatial hypersurfaces,
$$
{\dot{\cal R}}_H + 2 H {\cal R}_H \;=\;0\;\;\;\Rightarrow\;\;\; {\cal R}_H 
= {{\cal R}_H^0 \over a^2 } \;\;\;, \eqno(19c)
$$
and the averaged conservation law (14b) coincides with the local one,
$$
{\dot\varepsilon}_H + 3H (\varepsilon_H + p_H ) \;=\;0\;\;\;.\eqno(19d)
$$
For a given relation between $p_H$ and $\varepsilon_H$ the system of equations 
(19) is closed.
\bigskip
Alternatively, we may look at the homogeneous--isotropic models 
within the present framework in terms of a  time--dependent lapse function.
For the time variable $S$ and 
requiring $h = h_H (S)$, $\varepsilon = \varepsilon_H (S)$,
$p = p_H (S)$, ${\cal R} = {\cal R}_H (S)$, the domain dependence disappears,
and we have ${\tilde N}= h_H^{-1}$,
${\tilde \theta} = \theta_H (S) h_H^{-2}$, ${\tilde \theta}_H = 
3{\tilde H}$. Setting $\sigma = 0$ we also have ${\tilde{\cal Q}}_{\cal D}=0$,
but ${\tilde{\cal P}}_{\cal D}=-{h_H^{'} \over h_H}{\tilde\theta}_H \ne 0$.
The system of averaged
equations Eqs. (13) together with Eqs. (14) reduces to the following set: 
$$\eqalignno{
3{a'' \over a} &+ 4\pi G {1\over h_H^2}(\varepsilon_H + 3 p_H ) =
-3 {\tilde H} {h_H^{'} \over h_H}\;\;\;,&(20a)\cr
6{a'^2\over a^2} &+ {{\cal R}_H \over h_H^2} - 16\pi G 
{\varepsilon_H \over h_H^2}=0\;\;\;,&(20b)\cr
\left({{\cal R}_H \over h_H^2}\right)' + 2{\tilde H} 
{{\cal R}_H \over h_H^2} &= -2{h_H^{'} \over h_H}( 16\pi G  
{\varepsilon_H \over h_H^2} - 6 {\tilde H}^2 )\;\;\;,&(20c)\cr
\varrho_H^{'} + 3{\tilde H}\varrho_H &= 0 \;\;\;,&(20d)\cr}
$$
which, together with $h_H = {\varepsilon_H + p_H \over \varrho_H }$ and 
an equation of state $p_H = \alpha (\varepsilon_H )$, 
are four equations for the four unknown functions $a(S)$, $\varrho_H (S)$,
$\varepsilon_H (S)$, and ${\cal R}_H (S)$, but only three equations are independent.
We have to use the fact that, with $p_H = \alpha (\varepsilon_H )$, $\varepsilon_H$
can be expressed in terms of $\varrho_H$, which closes the system.

This set of equations
looks odd with respect to the hypersurfaces $S=const.$ However, upon
reinvoking the covariant time derivative, e.g., $\dot a = h_H a'$,
${\tilde H} = {\dot a \over a}h_H^{-1}$, $h_H$ in the first three equations 
disappears, and   
we recover the familiar form of Eqs.~(19), illustrating the covariance of the
averaged equations.  
\bigskip\bigskip
\noindent
{\rma 4.2. Inhomogeneous `Dust' Cosmologies}
\bigskip\medskip\noindent
Putting in Eqs. (13) all pressure terms to zero and noticing that 
$\varepsilon$ reduces to the restmass density we have $h\equiv 1$, the lapse 
function ${\tilde N}\equiv 1$, and the covariant time--derivative 
${d\over d\tau} = \partial_S$. Hence,
we directly recover the form of the averaged equations of Paper I
for cosmologies with a `dust' matter content:
$$
3{{\ddot a}_{\cal D} \over a_{\cal D}} + 4\pi G \langle \varrho\rangle_{\cal D}\;=\; {\cal Q}_{\cal D} 
\;\;\;;\eqno(21a)
$$
$$
6 H_{\cal D}^2 + \langle {\cal R} \rangle_{\cal D} 
- 16\pi G \langle \varrho \rangle_{\cal D} \;=\; -{\cal Q}_{\cal D} \;\;\;;\;\;\;
H:={{\dot a}_{\cal D} \over a_{\cal D}} \;\;\;,
\eqno(21b)
$$
with the integrability condition Eq. (14a) being
$$
{\cal Q}_{\cal D}^{\bdot} + 6 H_{\cal D} {\cal Q}_{\cal D} + \langle {\cal R} \rangle_{\cal D}^{\bdot}
+ 2 H_{\cal D} \langle {\cal R} \rangle_{\cal D} \;=\;
16\pi G (\langle\varrho\rangle_{\cal D}^{\bdot} + 
3 H_{\cal D} \langle\varrho\rangle_{\cal D}) \;\;\;;\eqno(21c)
$$
the balance equation (14d) reduces to the continuity equation for the 
averaged restmass density:
$$
\langle\varrho\rangle_{\cal D}^{\bdot} + 
3 H_{\cal D} \langle\varrho\rangle_{\cal D}\;=\;0\;\;\;.\eqno(21d)
$$
Notice that only in the case of an inhomogeneous `dust' model we can put the
lapse function $\tilde N$ or $N$ itself equal to $1$ 
without loss of generality; the averaged equations
are already covariant in the comoving and synchronous gauges.
It is to be emphasized that the `dust matter model' cannot generically be 
foliated into hypersurfaces $S=const.$ with an inhomogeneous lapse, 
for Eqs. (9) necessarily imply a 
constant lapse function for the geodesic condition of vanishing acceleration 
that itself is implied by vanishing pressure. 

\bigskip
Eqs. (21) form a set of four equations for the four unknown functions 
$a_{\cal D}$, $\langle \varrho \rangle_{\cal D}$, 
$\langle{\cal R}\rangle_{\cal D}$, and ${Q}_{\cal D}$, but only three equations
are independent. 
As discussed in Paper I, this system cannot be closed 
on the level of ordinary differential equations  unless additional 
(e.g. topological) constraints are imposed. Of course, this remark
also applies to the more general matter models. 
\bigskip
From Eqs. (21) it is obvious that the requirement ${\cal Q}_{\cal D} = 0$ is
necessary {\it and} sufficient in order that $a_{\cal D}(t)$ obeys the equations of
standard FLRW cosmologies.

\vfill\eject
\noindent
{\rma 4.3. Radiation--dominated Inhomogeneous Cosmologies}
\bigskip\medskip\noindent
Let us consider a situation in which radiation is directly coupled to the matter fluid
(according to the conjecture of a local thermodynamic equilibrium state of
radiation and matter) , then we may describe the radiation cosmos as a single 
component perfect fluid with 
radiation pressure $p_{\gamma}$ and radiation energy density 
$\varepsilon_{\gamma}$ obeying $\varepsilon_{\gamma} = 3 p_{\gamma}$ 
(Ellis 1971, see, however, Ehlers 1971). 
We infer already from the averaged conservation law (14b) that the time evolution 
of a radiation--dominated inhomogeneous
universe is different from that expected from the corresponding homogeneous--isotropic 
world model:
$$
\partial_t \langle\varepsilon_{\gamma}\rangle_{\cal D} + 4 {\tilde H}_{\cal D} 
\langle\varepsilon_{\gamma}\rangle_{\cal D} =
\langle {\partial_t p}_{\gamma}\rangle_{\cal D} - \partial_t 
\langle p_{\gamma} \rangle_{\cal D}\;\;\;.
$$
The term on the r.--h.--s. of this equation, which vanishes in the standard model, 
may be interpreted as an accumulated effect from inhomogeneities in the 
radiation field yielding  
deviations from a global `equilibrium equation of state'. 
It should be stressed that these deviations also occur  in the
case where the `backreaction' terms ${{\cal Q}}_{\cal D}$ and 
${{\cal P}}_{\cal D}$ are both found or assumed to be negligible. 
Therefore, radiation--dominated fluids deserve further detailed study.
\bigskip\bigskip
\noindent
{\rma 4.4. Inhomogeneous Scalar Field Cosmologies}
\bigskip\medskip\noindent
Following Bruni et al. (1992) we may describe the dynamics of a scalar field
$\phi$, minimally coupled to gravity, 
in terms of the natural slicing of spacetime
into a foliation of ${\phi = const.}$ hypersurfaces. 
Einstein's equations for a scalar field source are 
(under conditions stated below) equivalent to the 
phenomenological $3+1$--description of an evolving pressure--supported 
perfect fluid with 
energy--momentum tensor and corresponding fluid 4--velocity (normal to the hypersurfaces of constant 
$\phi$) (Taub 1973, Madsen 1988):
$$
T_{\mu\nu}^{\phi} = \varepsilon_{\phi} u_{\mu}u_{\nu} + p_{\phi} h_{\mu\nu}  
\;\;\;;\;\;\;u^{\mu} = {-\partial^{\mu}\phi \over \psi}\;\;\;.\eqno(22a)
$$ 
The magnitude $\psi$ normalizes the momentum density vector $\partial^{\mu}\phi$ so that $u^{\mu}u_{\mu} = -1$,
$$
0\; < \;\psi =\sqrt{-\partial^{\alpha}\phi \partial_{\alpha}\phi} = 
u^{\mu} \partial_{\mu}\phi = :{\dot\phi}\;\;\;,\eqno(22b)
$$ 
where the overdot stands for the material (or Lagrangian) derivative operator 
as before.
From Eq. (22a) we conclude that Einstein's equations feature the 
perfect fluid energy--momentum tensor (e.g., Madsen 1988):
$$
T_{\mu\nu}^{\phi} = 
(\varepsilon_{\phi}+p_{\phi}) u_{\mu}u_{\nu} + p_{\phi} g_{\mu\nu} \;=\;
\partial_{\mu}\phi \partial_{\nu}\phi - 
g_{\mu\nu}\left( {1\over 2}\partial^{\alpha}\phi \partial_{\alpha}\phi + 
V_{\rm eff}(\phi)\right)\;\;, 
\eqno(23a)
$$
with 
$$
\varepsilon_{\phi} = {1\over 2} \psi^2 + V_{\rm eff}
(\phi) 
\;\;\;,\eqno(23b)
$$ 
$$
p_{\phi} = {1\over 2} \psi^2 - V_{\rm eff} (\phi) 
\;\;\;.\eqno(23c)
$$ 
\smallskip\noindent
The fluid analogy is valid, if the fluid 4--velocity is time--like and, 
hence, the hypersurfaces 
$\phi = const.$ space--like. For this to be true 
the 4--gradient of the scalar field has to be time--like, 
$$
- \psi^2 = \partial_{\alpha}\phi\partial^{\alpha}\phi < 0 \;\;\;.\eqno(23d)
$$ 
This is a sufficient requirement for having the energy condition
$$
\varepsilon_{\phi} + p_{\phi} = \psi^2 > 0 \;\;\;,
\eqno(23e)
$$
which follows from Eqs. (23b,c).
This condition still allows for powerlaw inflation; exponential inflation
is excluded and has to be studied as a separate case. This case can be 
studied within the fluid analogy, if we model the constant effective 
potential with a cosmological constant. (The basic equations have to be 
used including the cosmological constant -- see Appendix). 

We finally note that, only for constant effective potential ($V_{\rm eff}' = 0$), 
the restmass conservation law of a perfect fluid corresponds to   
the Klein--Gordon equation 
$$
\dot\psi + \theta \psi + V_{\rm eff}'(\phi) = 0\;\;\;,\;\;\;\psi = {\dot\phi}
\;\;\;.\eqno(23f)
$$

\medskip\noindent
The `equation of state' of a scalar field is, in general, not barotropic 
(see: Bruni et al. 1992).
However, for interesting cases it is barotropic and 
can be represented in terms of an `equation of state'
$p_{\phi} = \alpha_{\phi} (\varepsilon_{\phi})$: 1) $\alpha_{\phi} = -1$ 
for a stationary state 
(vacuum ground state), and 2) $\alpha_{\phi}= +1$ 
for the free state (corresponding to a `stiff fluid').
In general, if there exists an equation of state, it will have the form 
$p_{\phi} = \alpha_{\phi} (\varepsilon_{\phi}, s_{\phi})$ 
with the entropy density $s_{\phi}$. However, an evolving scalar field
will in general yield a dependence of $p_{\phi}$ 
on the other dynamical variables $g_{\mu\nu}$ and $\phi$. 
The function $\alpha_{\phi}$ is determined by the dynamics and it may or may
not be a priori written, e.g., as a function of the density and the entropy density. 
\bigskip\noindent
Since the minimally coupled free scalar field (dilaton) is singled out in the
present investigation as the only inhomogeneous case in which the averaged 
equations attain a simpler form, it is certainly worth studying this case in more
detail. This is the headline of a forthcoming work (Buchert \& Veneziano 2001).

\bigskip\bigskip
\noindent
{\rma Acknowledgements:}
I would like to thank Gabriele Veneziano (CERN, Geneva),  
Ruth Durrer and Jean--Philippe Uzan (Univ. of Geneva), Mauro Carfora (Univ. of Pavia),
Toshifumi Futamase and Masahiro Takada (Univ. of Sendai),  Hideki Asada and Masumi
Kasai (Univ. of Hirosaki) for inspiring and helpful discussions. 
I am especially thankful to Gabriele Veneziano for his invitation to CERN,
where this work was prepared, and to Ruth Durrer for her invitation to
Geneva University, where it was completed during visits in 1999 and 2000 
with support by the Tomalla Foundation, Switzerland.

\vfill\eject

\noindent{\rmm Appendix:}
\smallskip\noindent
{\rmm Basic Equations in the ADM Formalism}
\bigskip\bigskip
\noindent
Let $n_{\mu}$ be the future directed unit normal to the hypersurface $\Sigma$.
The projector into $\Sigma$, $h_{\mu\nu} = g_{\mu\nu} + n_{\mu}n_{\nu}\;,\;\
(\Rightarrow h_{\mu\nu}n^{\mu} = 0\;,\;h^{\mu}_{\;\,\nu}h^{\nu}_{\;\,\gamma}=
h^{\mu}_{\;\,\gamma})$, induces in $\Sigma$ the 3--metric 
$$
h_{ij}: = g_{\mu\nu}h^{\mu}_{\;\,i}h^{\nu}_{\;\,j}\;\;\;.\eqno(A1a)
$$
Let us write
$$
n_{\mu} = N (-1,0,0,0)\;\;\;\;\;\;,\;\;\;\;\;\;n^{\mu} = {1\over N}(1,-N^i )
\;\;\;,\eqno(A1b)
$$ 
with the lapse function $N$ and the shift vector $N^i$. Note that 
$N$ and $N^i$
can be determined by the choice of coordinates.

\noindent 
From $n_{\mu} = g_{\mu\nu}n^{\nu}$ we find $g_{00}=-(N^2 - N_i N^i )$; $g_{0i}=N_i$; $g_{ij}=h_{ij}$ and,
setting $x^0 = t$, the line element becomes:
$$
ds^2 = - (N^2 - N_i N^i )dt^2 + 2N_i dt dx^i + g_{ij}dx^i dx^j = 
-N^2 dt^2 + g_{ij} (dx^i + N^i dt)(dx^j + N^j dt) \;\;\;.\eqno(A1c)
$$
Introducing the extrinsic curvature tensor on $\Sigma$ by
$$
K_{ij} : = -n_{\mu;\nu}h^{\mu}_{\;\,i}h^{\nu}_{\;\,j} = -n_{i;j}\;\;\;,\eqno(A1d)
$$
we obtain the ADM equations (Arnowitt et al. 1962, York 1979):

\bigskip\medskip\noindent
\underbar{Energy (Hamiltonian) constraint:}
$$
{\cal R} - K^i_{\;\,j}K^j_{\;\,i} + K^2 = 16\pi G \varepsilon + 2\Lambda\;\;\;,\;\;\;
\varepsilon: = T_{\mu\nu}n^{\mu}n^{\nu}\;\;\;;\eqno(A2a)
$$
\smallskip\noindent
\underbar{Momentum constraints:}
$$
K^i_{\;\,j||i} - K_{||j} = 8\pi G J_j \;\;\;,\;\;\;J_i : = - T_{\mu\nu}n^{\mu}h^{\nu}_{\;\,i}\;\;\;;\eqno(A2b)
$$
\smallskip\noindent
\underbar{Evolution equation for the first fundamental form:}
$$
{1\over N}\partial_t g_{ij} = -2 K_{ij} + {1\over N} (N_{i||j} + N_{j||i}) \;\;\;;\eqno(A2c)
$$
\smallskip\noindent
\underbar{Evolution equation for the second fundamental form:}
$$
{1\over N} \partial_t K^i_{\;\,j} = {\cal R}^i_{\;\,j} + 
K K^i_{\;\,j} - \delta^i_{\;\,j} \Lambda 
- {1\over N}{N^{||i}}_{||j} + {1\over N} \left(
K^i_{\;\,k}N^k_{\;\,||j} - K^k_{\;\,j}N^i_{\;\,||k} + N^k K^i_{\;\,j||k}
\right)
$$
$$
- 8\pi G  ( {\cal S}^i_{\;\,j} + {1\over 2}\delta^i_{\;\,j} (\varepsilon - {\cal S}^k_{\;\,k}))\;\;\;,\eqno(A2d)
$$
where ${\cal S}_{ij} : = T_{\mu\nu} h^{\mu}_{\;\,i}h^{\nu}_{\;\,j}\;$.
\bigskip
\noindent
For the trace parts of (A2c) and (A2d) we have:
\smallskip
$$
{1\over N} \partial_t g = 2g (-K + {1\over N}N^k_{\;\,||k})\;\;\;,\;\;\;g:=\det (g_{ij} )\;\;\;;\eqno(A2e)
$$
$$
{1\over N} \partial_t K =  {\cal R} + K^2 - 4\pi G (3\varepsilon - {\cal S}^k_{\;\,k} ) - 3\Lambda  
- {1\over N}{N^{||k}}_{||k} + {1\over N} N^k K_{||k}\;\;\;.\eqno(A2f) 
$$
\bigskip\noindent
For our purpose of averaging we have used equations that correspond to the coordinate choice
of vanishing shift vector. Thus, all inhomogeneities of the fluid were put into the 3--metric and the lapse function.

\medskip\noindent
Assuming the tensor $T_{\mu\nu}$ has the form $T_{\mu\nu} = \varepsilon u_{\mu}u_{\nu} + p h_{\mu\nu}$ and
putting the shift vector $N^i = 0$ and also $\Lambda = 0$, we obtain the equations of the main text by
defining the lapse function in such a way that 
$a_i ={N_{||i}\over N}\equiv {- p_{||i}\over \varepsilon + p} =
-{h_{||i}\over h}$.

\smallskip\noindent
Notice that with this choice the unit normal coincides with the 4--velocity and, especially, the momentum flux
density in $\Sigma$ vanishes. The total time--derivative operator
of a tensor field $\cal F$ along integral curves
of the unit normal, ${d\over d\tau}{\cal F}: = n^{\nu}\partial_{\nu}{\cal F}
= u^{\nu}\partial_{\nu}{\cal F}$ becomes 
$$
{d\over d\tau}{\cal F} = {1\over N}\partial_t  {\cal F}\;\;\;,\eqno(A3a)
$$
since $n^{\nu}{\cal F}_{||\nu} = 0$.
Note that, although the definition of proper time is $\tau: = \int{N dt}$,
the line element cannot be written in the form of the comoving gauge by
measuring ``time'' through proper time $d\tau = N dt$, 
since $d\tau$ is not an exact form in the case of an inhomogeneous
lapse function. The exterior derivative of the proper
time will involve a non--vanishing shift vector according to the 
space--dependence of the lapse function. Therefore, 
a foliation into hypersurfaces $\tau = const.$
with simultaneously requiring $u_{\alpha} = - \partial_{\alpha}\tau$ is not
possible.

\medskip

For vanishing shift vector the line element reads:
$$
ds^2 = -N^2 dt^2 + g_{ij}dX^i dX^j \;\;\;.\eqno(A3b)
$$
The lapse function itself may be written explicitly: 
from $h = \dot S$ we have:
$$
N = {1\over h} \partial_t S \;\;\;.\eqno(A3c)
$$
Note that, if we assume that $h > 0$, implying the energy condition 
$\varepsilon + p > 0$, the proper time advances only 
in periods when the derivative $\partial_t S  > 0$ which 
makes a difference if we consider fluids that mimick a scalar field source.

\noindent  
The coordinates in $\Sigma$ are written in capital letters now, because for vanishing shift vector they correspond
to Lagrangian coordinates as in classical fluid mechanics. 
In these coordinates ${d\over d\tau} = {1\over N}{d\over dt} =
{1\over N} \partial_t \vert_{X^i}$.


\vfill\eject

\centerline{\rmm References}
\bigskip\bigskip
\ref
Adler S., Buchert T. (1999): {\it Astron. Astrophys.} {\bf 343}, 317.
\ref
Arnowitt R., Deser S., Misner C.W. (1962): in {\it Gravitation: an Introduction to Current Research}, L. Witten (ed.),
New York: Wiley
\ref
Bruni M., Dunsby P.K.S., Ellis G.F.R. (1992a): {\it Ap.J.} {\bf 395}, 34.
\ref
Bruni M., Ellis G.F.R., Dunsby P.K.S. (1992b): {\it Class. Quant. Grav.} {\bf 9}, 921
\ref
Buchert T. (2000): {\it G.R.G.} {\bf 32}, 105. (Paper I).
\ref
Buchert T., Dom\'\i nguez A. (1998): {\it Astron. Astrophys.} {\bf 335}, 395.
\ref
Buchert T., Dom\'\i nguez A., P\'erez--Mercader J. (1999): {\it Astron. Astrophys.}
{\bf 349}, 343.
\ref
Buchert T., Ehlers J. (1997): {\it Astron. Astrophys.} {\bf 320}, 1.
\ref
Buchert T., Veneziano G. (2001): in preparation.
\ref
Dunsby P.K.S., Bruni M., Ellis G.F.R. (1992): {\it Ap.J.} {\bf 395}, 54. 
\ref
Ehlers J. (1961): {\it Akad. Wiss. Lit. (Mainz); Abh. Math.--Nat. Kl.} 
{\bf No. 11}, 793; translation: {\it G.R.G.} {\bf 25}, 1225 (1993). 
\ref
Ehlers J. (1971): in ``General Relativity and Cosmology'', Proc. XLVII Enrico Fermi
School, R.K. Sachs (ed.), New York: Academic, pp.1--67.
\ref
Ellis G.F.R. (1971): in ``General Relativity and Cosmology'', Proc. XLVII Enrico Fermi
School, R.K. Sachs (ed.), New York: Academic, pp.104--179.
\ref
Ellis G.F.R., Bruni M. (1989): {\it Phys. Rev. D} {\bf 40}, 1804
\ref
Ellis G.F.R., Bruni M., Hwang J. (1990): {\it Phys. Rev. D.} {\bf 42}, 1035
\ref
Hwang J., Vishniac E. (1990): {\it Ap.J.} {\bf 353}, 1
\ref
Israel W. (1976): {\it Ann. Phys.} {\bf 100}, 310.
\ref
Kasai M. (1995): {\it Phys. Rev. D} {\bf 52}, 5605.
\ref
King A.R., Ellis G.F.R. (1973): {\it Comm. Math. Phys.} {\bf 31}, 209
\ref
MacCallum M.A.H., Taub A.H. (1972): {\it Comm. Math. Phys.} {\bf 25}, 173
\ref
Madsen M.S. (1988): {\it Class. Quant. Grav.} {\bf 5}, 627
\ref
Maartens R., Triginer J., Matravers D.R. (1999): {\it Phys. Rev. D} {\bf 60}, 103503.
\ref
Stoeger W.R., Helmi A., Torres D.F. (1999): gr--qc/9904020
\ref
Takada M., Futamase T. (1999): {\it G.R.G.} {\bf 31}, 461.
\ref
Taub A.H. (1973): {\it Comm. Math. Phys.} {\bf 29}, 79
\ref
Yodzis P. (1974): {\it Proc. Royal Irish Acad.} {\bf 74A}, 61.
\ref
York J.W. Jr. (1979): in ``Sources of Gravitational Radiation'', L. Smarr (ed.), 
Cambridge Univ. Press, p.83 

\vfill\eject
\bye